

\magnification=\magstep1
\vsize=23.5true cm
\hsize=16true cm
\topskip=1true cm
\abovedisplayskip=3mm
\belowdisplayskip=3mm
\abovedisplayshortskip=0mm
\belowdisplayshortskip=2mm
\normalbaselineskip=12pt
\normalbaselines
\def\Dbar{\hbox{{\it D}\hskip-4.5pt/}}


\def\vsp{\vskip 0.4truecm \par}
\def\ts{\thinspace}

\def\ni{\noindent}

\def\rp{\item{}}
\def \r#1{[\ts {#1}\ts \ts ]} \def\rf#1{\item{\r{#1} \ }}

\centerline{\bf ANOMALY AND THERMODYNAMICS FOR 2D SPINORS}
\centerline{\bf IN THE AHARONOV--BOHM GAUGE FIELD }

\medskip

\centerline{ROBERTO  SOLDATI}
\centerline{{\it Dipartimento di Fisica  ``Augusto Righi'',
                 sezione I.N.F.N. Bologna}}
\centerline{{\it via Irnerio 46 - 40126 Bologna (Italia)}}

\smallskip
\centerline{and}
\smallskip
\centerline{PAOLA GIACCONI, FABIO MALTONI}
\centerline{{\it Dipartimento di Fisica  ``Augusto Righi'',
                 sezione I.N.F.N. Bologna}}
\centerline{{\it via Irnerio 46 - 40126 Bologna (Italia)}}

\medskip

\centerline{ABSTRACT}
\smallskip
\noindent
{\it The axial anomaly is computed for Euclidean Dirac fermions on the plane.
The dependence upon the self-adjoint extensions of the Dirac operator is
investigated and its relevance concerning the second
virial coefficients of the anyon gas is discussed}.
\medskip
\noindent
{\bf 1.\quad Spinors on the plane: A-B statistical interaction}
\smallskip
The coupling of two dimensional fermions with the Aharonov--Bohm
(A-B) gauge field leads to statistics transmutation and then to spinning
anyon matter, whose special interesting features are worth being considered.
\par
We first study the Euclidean Dirac operator on the plane
$$
i\Dbar=i\gamma_{\mu} (\partial_{\mu} - ieA_{\mu})\quad,
\eqno{(1.1)}
$$
\noindent
the (A-B) gauge potential being
$A_{\mu} \equiv{\alpha \over e}
\epsilon_{\mu\nu} {x_{\nu}\over x_{1}^{2}+x_{2}^{2}}$,
where $\gamma_1=\sigma_1,\ \ \gamma_2=\sigma_2,\ \ \gamma_3=i\gamma_1\gamma_2=
\sigma_3$, {\it sigma}'s  being the Pauli matrices,
$\epsilon_{12}=1$ and $-1<\alpha<0$.
As  is well known the field strength is $F_{\mu\nu}=
-{2\pi\alpha \over e} \epsilon_{\mu\nu} \delta^{(2)}(x)$.\par
Now let us consider the eigenvalues
and eigenfunctions of the Dirac operator . The crucial point is that,
in order to find a
{\underbar {complete}}
orthonormal basis which diagonalizes the Dirac operator it is necessary
to consider
its self-adjoint extensions [1][2].
To this concern let us choose polar coordinates $(r,\phi)$ on the plane
and rescale the spinor wave function as
$$
{1\over \sqrt \mu}\psi_\lambda(r,\phi)\longmapsto
\psi_{\lambda,n}(\xi,\phi)\equiv
\left|\matrix{\psi_\lambda^{(L)}(\xi)e^{in\phi}\cr
              \psi_\lambda^{(R)}(\xi)e^{i(n+1)\phi} \cr}\right|\quad,
\eqno{(1.2)}
$$
\noindent
where $\mu$ is a suitable mass parameter to fix the scale of the eigenvalues
(a natural choice is to set $\mu=e$), $n\in {\bf Z},\ \ \lambda
\in {\bf R}-\{0\}$ and $\xi=\mu r$. When $n\not=0$
we get the eigenspinors regular at the origin: namely,
$$
\psi_{\lambda,\pm n}(\xi,\phi)= {\sqrt {|\lambda| \over 4\pi}}
\left|\matrix{(\pm i) J_{\pm\nu}(|\lambda|\xi)e^{\pm in\phi} \cr
              sgn(\lambda)J_{\pm (\nu+1)}(|\lambda|\xi)e^{i(1 \pm n)\phi}
\cr}\right|\quad;
\eqno(1.3)
$$
\noindent
here $n \in {\bf N}$, $J_\nu$ being the Bessel function of order
$\nu(\pm n) \equiv \pm n+\alpha $.
On the other hand, the partial waves corresponding to $\nu(0) \equiv \alpha $
can not be both regular at the origin unless completeness of the
eigenfunctions is lost [1]. Then one has to consider
the self-adjoint extensions  of the Dirac operator by means of
the standard
Von Neumann method of the deficiency indices.
The corresponding eigenfunctions for $ \nu=\alpha$ can be written in the form

$$
\eqalign
{&\psi_{\lambda,0}^{(\omega)}(\xi,\phi)=
{\sqrt {|\lambda|\over 4\pi(1+\sin\theta(|\lambda|)\cos\alpha\pi)}}\ \times\cr
&\left|\matrix{i\cos {\theta(|\lambda|) \over 2} J_\alpha (|\lambda|\xi)-
              i\sin {\theta(|\lambda|) \over 2} J_{-\alpha}(|\lambda|\xi) \cr
 sgn(\lambda) \left[\cos {\theta(|\lambda|) \over 2}
J_{(1+\alpha)}(|\lambda|\xi) +
               \sin {\theta(|\lambda|) \over 2}
J_{-(1+\alpha)}(|\lambda|\xi)\right]
               e^{i\phi}\cr}\right|\cr}\quad,
\eqno(1.4)
$$
where
$$
\tan\theta(|\lambda|) = |\lambda|^{2\alpha+1}\tan\omega\quad .
\eqno(1.5)
$$
\noindent
The eigenfunctions in eq.s~(1.3),(1.4) are improper eigenfunctions, since they
belong to eigenvalues of the continuous spectrum. They are suitably
normalized according to theory of the distributions, $viz.$

$$
\lim_{R\to\infty}\int_0^{\mu R}\xi d\xi\int_0^{2\pi}d\phi\
\psi^{\dag}_{n_1}(|\lambda_1|\xi,\phi)\psi_{n_2}(|\lambda_2|\xi,\phi)=
\delta_{n_1 n_2}\delta(\lambda_1-\lambda_2)\quad .
\eqno(1.6)
$$

Moreover, in order to obtain the correct normalization as in eq.~(1.6),
one has to put the contribution at the origin equal to zero,
thereby finding the relationship of eq.~(1.5). For any value of $\omega$,
the  purely continuous spectrum is
the whole real line, due to the absence of zero modes.
\smallskip
\noindent
{\bf 2.\quad The axial anomaly on the plane}
\smallskip
Once the eigenvalue problem has been solved, we are able
to set up the complex power by means of the spectral theorem.
The complex power of the dimensionless operator
$I_{\omega}^{-s}\equiv \left({\Dbar_\omega \over \mu}\right)^{-s}$
is defined by the  kernel
$$
\eqalign{
& <\xi_1,\phi_1|I_\omega^{-s}|\xi_2, \phi_2> \equiv
  K_{-s}(I_\omega;\xi_1,\phi_1,\xi_2,\phi_2)\cr
& = 2\ \sum_{n=1}^{\infty} \left [\int_0^{\infty} d\lambda\
\lambda^{-s} \psi_{\lambda,n}(\xi_1,\phi_1)
\psi_{\lambda,n}^{\dag}(\xi_2,\phi_2) + (n \rightarrow -n)\right ]\cr
& +2\ \int_0^{\infty} d\lambda
\ \lambda^{-s} \psi_{\lambda,0}^{(\omega)}
(\xi_1,\phi_1)\psi_{\lambda,0}^{(\omega)\dag}(\xi_2,\phi_2)\quad , \cr}
\eqno{(2.1)}
$$
which can be analytically extended to a meromorphic function of the complex
variable $s$; the key property is that the kernel of the complex power is
regular at $s=0$.
In particular, the value of its trace over spinor
indices, on the diagonal $(\xi_1,\phi_1)=(\xi_2,\phi_2)$,
can be explicitely evaluated [3] either in the case
$\omega=0,\pi,\quad  -1<\alpha<0$, or in the case of semions: namely,
$\alpha= -{1\over 2}, \omega\in {\bf R}$.
We can properly construct
the euclidean averages of the vector and axial currents, respectively,
by means of point--splitting as well as analytic continuation [4], namely
$$
<j_\mu^{(\omega)}(x)>=e<tr[\gamma_\mu\psi(x)\psi^{\dag}(x)]>\equiv
\lim_{s\to 1}\lim_{\epsilon\to 0}e\
tr[\gamma_\mu K_{-s}(I_\omega;x,x+\epsilon)]
\eqno (2.2a)
$$
where $<\cdot>$ means euclidean average and
$$
<j_{\mu 3}^{(\omega)}(x)>=e<tr[\gamma_\mu\gamma_3\psi(x)\psi^{\dag}(x)]>
\equiv
\lim_{s\to 1}\lim_{\epsilon\to 0}e\ tr[\gamma_\mu\gamma_3
K_{-s}(I_\omega;x,x+\epsilon)]
\eqno (2.2b)
$$
{}From the above definitions of the averaged local currents
it is straightforward to show the quantum balance equations: namely,\quad
$<\partial_\mu j_\mu^{(\omega)}(x)> = 0$,
testing the gauge invariance of the definition in eq.~(2.2a), whereas
$$
<\partial_\mu j_{\mu 3}^{(\omega)}(x)> = 2ie\ \lim_{s\to 0}
\ tr[\gamma_3\
K_{-s}(I_\omega;x,x)]\equiv {\cal A}^{(\omega)}(x)\quad
\eqno (2.3)
$$
leads to the definition of the local axial anomaly, once the topology has
been chosen in taking the limit $s\to 0$; we shall discuss below this delicate
matter. \par

A first possibility is to consider the ${\cal S}^\prime$-topology.
If we take the limit $s\to 0$ in the sense of the distributions, it is
straightforward to show that the limit exists only for $\omega=0,\pi$ and we
get
$$
\int d^2 x\  {\cal A}^{(0,\pi)}(x)f(x)= -\alpha\quad ,
\eqno(2.4)
$$
where $f$ is a suitable test function belonging to ${\cal S}({\bf R}^2)$
normalized to $f(0)=\mu$; we notice that the above result, in
full agreement with the one of Ref.[5], actually corresponds to the usual
formula, $viz.$ ${\cal A}^{(0,\pi)}(x)=-{ie^2\over 2\pi}\epsilon_{\mu\nu}
F_{\mu\nu}(x)$ as a distribution.\par
A second possibility is to consider the limit $s\to 0$ in the natural
topology of ${\bf R}-\{0\}$ and {\underbar {only afterwards}}
continue the result to
${\cal S}^\prime ({\bf R}^2)$. As a matter of fact, a non vanishing result is
obtained in this case for $\omega\not= 0,\pi$ and, when $\alpha=-{1\over 2}$,
we
can compute explicitely
$$
\lim_{s\to 0}\lim_{\epsilon\to 0}e\ tr[\gamma_3
K_{-s}(I_\omega;x,x+\epsilon)]|_{\alpha=-{1\over 2}}
={ie\sin\omega\over 2\pi^2r^2},\quad r\not= 0\quad .
\eqno(2.5)
$$
As a consequence, there is a unique continuation in ${\cal S}^\prime({\bf
R}^2)$
which reads ($\alpha=-{1\over 2}$)
$$
\int_0^{\infty} \xi d\xi \int_0^{2\pi}d{\phi}\ f(\xi,\phi){\cal A}^{(\omega)}
(\xi)= \int_0^{\infty} \xi d\xi \int_0^{2\pi}d{\phi}\ {ie\sin\omega\over
2\pi^2[\xi^2]}f(\xi,\phi)\quad ,
\eqno(2.6)
$$
where we recall the definition
$$
{1\over [\xi^2]}\equiv {1\over 2\mu^2}
(\partial_1^2 + \partial_2^2)(\ln r)^2 + C(\mu)\delta^{(2)}(x)\quad ,
\eqno (2.7)
$$
the arbitrary function $C(\mu)$
being  there in order to guarantee the scaling property
${1\over [\xi^2]} = {1\over \mu^2}\cdot {1\over [r^2]}$.
We stress that, in the present case,
 if the test function vanishes at the origin, where
 the field strength is concentrated, still a nonvanishing contribution
survives, of a purely quantum mechanical nature, which depends upon the
parameter of the self--adjoint extensions, a quite interesting feature
closely reminescent of the AB effect.\par

\smallskip
\noindent
{\bf 3.\quad Thermodynamics and anomaly}
\smallskip
The knowledge of the eigenvalues and eigenfunctions of $\Dbar_{\omega}$
allows to compute the 2nd virial coefficient for a gas of spinning anyons
in 2+1 dimensions: namely,
$$
a_2=\mp{\pi\beta\over 2m}\left(1\pm 4\int d^2x \ [G_{int}(\beta;x,x) \pm
G_{int}(\beta;x,-x)]\right)\quad ,
\eqno (3.1)
$$
where the upper and lower signs refer to bosons and fermions respectively,
with
$$
G_{int}(\beta;x,y)= tr <x|e^{-\beta H(\alpha)} - e^{-\beta H(0)}|y>\quad ;
\eqno (3.2)
$$
here $H(\alpha)= \Dbar_{\omega}^2$ is the 2-body spinning relative Hamiltonian
while $tr$ means trace over spinor indices.
Since we are on the plane, eq.~(3.1) needs some volume regularization and, in
the present case, the suitable one is dimensional regularization
to deal with products of regular and singular wavefunctions.
The explicit calculation in the case
$\omega = 0,\pi$ gives
$$
a_2= \mp{\pi\beta\over 2m}(1\pm 2\alpha^2)\quad .
\eqno (3.3)
$$
It turns out that the above result is different (for bosons) from the one of
the standard spinless case [6] and, consequently, it appears that spin and
singular wavefunctions indeed carry new features into the anyon physics.
A further interesting point is the existence of a nice relationship
between the 2nd virial coefficient and the axial anomaly. In the present case
$\omega=0,\pi$ one obtains
$$
{m\over 2\pi\beta}[a_2^F(\alpha+1) - a_2^B(\alpha)]=
\int d^2 x\  {\cal A}^{(0,\pi)}(x)f(x)\quad ,
$$
where $f(0)=m$. The corresponding relationships can be also found in the semion
case for arbitrary values of $\omega$. It would be very interesting to
understand the properties of the N-anyon spinning gas and, in particular,
the possible role of the singularities in the structure of the ground state.

\vsp
\ni{{\bf  References}} \vsp

\rf1  P. de Sousa Gerbert, Phys. Rev. {\bf D 40} (1989) 1346;
\rp   P. de Sousa Gerbert and R. Jackiw,
      Comm. Math. Phys. {\bf 124} (1989) 229;
\rf2  M. Alford, J. March-Russell and F. Wilczek, Nucl. Phys. {\bf B328} (1989)
      140;

\rp   C. Manuel and R. Tarrach, Phys. Lett. {\bf 268B} (1991) 222 and Phys.
      Lett. {\bf 301B} (1993) 72;
\rp   J. Grundberg, T.H. Hansson, A. Karlhede and J. M. Leinaas,
      Mod. Phys. Lett. {\bf B} (1991) 539.
\rf3  P. Giacconi, S. Ouvry and R. Soldati, Bologna preprint DFUB/94-13,
      to appear in Phys. Rev. D.
\rf4  M. Hortacsu, K.D. Rothe and B. Schroer, Phys. Rev. {\bf D 20}
      (1979) 3023.
\rf5  A. Comtet and S. Ouvry, Phys. Lett. {\bf 225B} (1989) 272.
\rf6  D. P. Arovas, R. Schrieffer, F. Wilczek and A. Zee, Nucl. Phys. {\bf
      B251} (1985) 117.
\vfill\eject
\bye